\documentclass[conference]{IEEEtran}
\usepackage[utf8]{inputenc} 
\usepackage[T1]{fontenc}    
\usepackage{hyperref}       
\usepackage{url}            
\usepackage{booktabs}       
\usepackage{amsfonts}       
\usepackage{nicefrac}       
\usepackage{microtype}      
\usepackage{xcolor}         

\usepackage{latexsym}

\usepackage{amsmath,amsfonts,bm}

\def\eqref#1{equation~\ref{#1}}

\def\1{\bm{1}}

\def\rvc{{\mathbf{c}}}

\def\rve{{\mathbf{e}}}

\def\rvt{{\mathbf{t}}}

\def\rvx{{\mathbf{x}}}
\def\rvy{{\mathbf{y}}}
\def\rvz{{\mathbf{z}}}

\def\rmC{{\mathbf{C}}}

\def\rmT{{\mathbf{T}}}

\def\rmX{{\mathbf{X}}}
\def\rmY{{\mathbf{Y}}}
\def\rmZ{{\mathbf{Z}}}

\def\mI{{\bm{I}}}

\DeclareMathAlphabet{\mathsfit}{\encodingdefault}{\sfdefault}{m}{sl}
\SetMathAlphabet{\mathsfit}{bold}{\encodingdefault}{\sfdefault}{bx}{n}

\usepackage{titletoc}
\usepackage{wrapfig} 

\newcommand{\minimize}{%
  \mathopen{}\operatorname*{minimize}%
}

\usepackage{tabularx}
\usepackage{graphicx}
\usepackage{algorithm}
\usepackage{graphicx}
\usepackage{algpseudocode}
\algtext*{EndWhile}
\algtext*{EndIf}
\algtext*{EndFor}
\usepackage{lipsum}
\usepackage{amsmath}
\usepackage[T1]{fontenc}
\makeatletter
\newcommand*{\algrule}[1][\algorithmicindent]{%
  \hspace*{.2em}
  \vrule 
  \hspace*{\dimexpr#1-.2em-.4pt}%
}

\newcommand{\StatePar}[1]{%
  \State\parbox[t]{\dimexpr\linewidth-\ALG@thistlm}{\strut #1\strut}%
}
\renewcommand{\ALG@beginalgorithmic}{\offinterlineskip}

\newcount\ALG@printindent@tempcnta
\def\ALG@printindent{%
  \ifnum \theALG@nested > 0
    \ifx\ALG@text\ALG@x@notext
    \else
      \unskip
      \ALG@printindent@tempcnta=1
      \loop
        \algrule[\csname ALG@ind@\the\ALG@printindent@tempcnta\endcsname]%
        \advance \ALG@printindent@tempcnta 1
        \ifnum \ALG@printindent@tempcnta<\numexpr\theALG@nested+1\relax
      \repeat
        \fi
    \fi
}
\patchcmd{\ALG@doentity}{\noindent\hskip\ALG@tlm}{\ALG@printindent}{}{\errmessage{failed to patch}}
\makeatother

\usepackage{subfloat}

\algrenewcommand\algorithmicend{\strut\textbf{end}}
\algrenewcommand\algorithmicdo{\strut\textbf{do}}
\algrenewcommand\algorithmicwhile{\strut\textbf{while}}
\algrenewcommand\algorithmicfor{\strut\textbf{for}}
\algrenewcommand\algorithmicforall{\strut\textbf{for all}}
\algrenewcommand\algorithmicloop{\strut\textbf{loop}}
\algrenewcommand\algorithmicrepeat{\strut\textbf{repeat}}
\algrenewcommand\algorithmicuntil{\strut\textbf{until}}
\algrenewcommand\algorithmicprocedure{\strut\textbf{procedure}}
\algrenewcommand\algorithmicfunction{\strut\textbf{function}}
\algrenewcommand\algorithmicif{\strut\textbf{if}}
\algrenewcommand\algorithmicthen{\strut\textbf{then}}
\algrenewcommand\algorithmicelse{\strut\textbf{else}}

\algrenewcommand\algorithmicrequire{\strut\textbf{Input:}}
\algrenewcommand\algorithmicensure{\strut\textbf{Output:}}

\let\oldState\State
\renewcommand{\State}{\oldState\strut}

\usepackage{caption}
\usepackage{subcaption}
\usepackage{amsmath}
\usepackage{amssymb}
\usepackage{mathtools}
\usepackage{amsthm}

\usepackage[capitalize,noabbrev]{cleveref}

\theoremstyle{plain}

\theoremstyle{definition}
\theoremstyle{definition}

\IEEEoverridecommandlockouts
\begin{document} 
\title{\huge{Generative Diffusion Model-based Compression of MIMO CSI}}
\author{Heasung Kim$^{1}$\thanks{Part of this work was done during an internship at InterDigital Communications.}, Taekyun Lee$^{1}$, Hyeji Kim$^1$, Gustavo De Veciana$^1$, \\ Mohamed Amine Arfaoui$^2$, Asil Koc$^2$, Phil Pietraski$^2$, Guodong Zhang$^2$, and John Kaewell$^2$.\\
\centerline{$^1$The University of Texas at Austin \quad\quad $^2$InterDigital Communications \quad\quad}\\
Emails: {heasung.kim@utexas.edu, taekyun@utexas.edu, hyeji.kim@austin.utexas.edu, deveciana@utexas.edu} \\ 
\{MohamedAmine.Arfaoui, asil.koc, Philip.Pietraski, guodong.zhang, john.kaewell\}@interdigital.com
}
\maketitle 
\begin{abstract}
While neural lossy compression techniques have markedly advanced the efficiency of Channel State Information (CSI) compression and reconstruction for feedback in MIMO communications, efficient algorithms for more challenging and practical tasks—such as CSI compression for future channel prediction and reconstruction with relevant side information—remain underexplored, often resulting in suboptimal performance when existing methods are extended to these scenarios. To that end, we propose a novel framework for compression with side information, featuring an encoding process with fixed-rate compression using a trainable codebook for codeword quantization, and a decoding procedure modeled as a backward diffusion process conditioned on both the codeword and the side information. Experimental results show that our method significantly outperforms existing CSI compression algorithms, often yielding over twofold performance improvement by achieving comparable distortion at less than half the data rate of competing methods in certain scenarios. These findings underscore the potential of diffusion-based compression for practical deployment in communication systems.
\end{abstract}

\section{Introduction}
\label{section_introduction}
In Multiple-Input Multiple-Output (MIMO) communications, data transmission efficiency is closely tied to timely and accurate Channel State Information (CSI) feedback from a User Equipment (UE) to a Base Station (BS) \cite{guo2022overview}. However, with CSI often comprising hundreds to thousands of elements \cite{dahlman20134g, zaidi20185g} in modern communications, managing the feedback load becomes challenging. To address this, \emph{lossy compression} techniques have been explored to compress CSI and reduce feedback overhead while retaining critical signal information.

Recent advancements in neural network-based compression approaches  \cite{balle2016end, balle2018variational, theis2022lossy_compressive}, commonly referred to as \emph{neural lossy compression} \cite{yang2023introduction}, have demonstrated substantial improvements in compression performance over conventional techniques, such as JPEG \cite{wallace1991jpeg} and Vector Quantization \cite{ziv1985universal} for various compression tasks including CSI compression \cite{guo2022overview}.
The success of neural compression methods is largely attributed to their data-driven nature and ability to learn more effective transformation modules \cite{minnen2018joint}.

A pioneering application of neural lossy compression for CSI compression \cite{csinet_ver1} employs convolutional neural networks to achieve significant gains over conventional compressive sensing methods. Following this, several enhanced neural network architectures have been proposed for the task, including \cite{csinet_ver2, li2020novel, liu2021hyperrnn, lu2020crnet}. 
Recent research has focused on advanced coding schemes such as entropy coding or multi-rate coding for CSI compression \cite{park2024multi, ravula2021deep, nerini2022machine, kim2022learning}.
Despite these advancements, achieving further improvements in CSI compression remains a challenge. The need for enhanced performance is critical to advancing the reliability and efficiency of data transmission in next-generation communication systems, a topic actively discussed in standardization bodies \cite{9927255}.

Recently, further advancement in neural lossy compression has been achieved through the application of diffusion models \cite{yang2023diffusion}, particularly those employing U-Net architectures \cite{ronneberger2015unet}, which have gained recognition for their exceptional performance in image generation and restoration tasks \cite{croitoru2023diffusion}. The representational and reconstructive capabilities of these models have encouraged researchers to explore their potential in a variety of compression tasks. Notable efforts have focused on integrating diffusion models into compression pipelines, such as transmitting corrupted versions of the input source to the receiver for decoding with a diffusion model \cite{theis2022lossy}, or designing codeword-conditioned diffusion-based decoding processes \cite{yang2024lossy}.

\begin{figure}[t]
\begin{center}
\centerline{\includegraphics[width=0.33\textwidth]{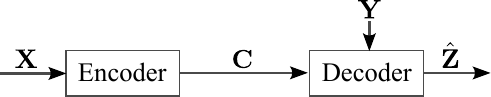}}
\caption{{System Model (Coding for computing with side information)}}
\label{fig_system_model}
\end{center}
\vspace{-0.4in}
\end{figure}
Existing work on diffusion models for compression has predominantly concentrated on natural image compression, with the primary objective of generating perceptually realistic reconstructions. While this has addressed important challenges in visual data processing, it does not capture a broader set of industrial applications where the objective extends beyond image reconstruction to computing specific target functions from compressed data.

In this paper, we explore the application of diffusion models to the \emph{Wyner-Ziv coding} type problems, i.e., compression with side information \cite{rebollo2005generalization}, where an encoder compresses an input source $\rmX$ into a codeword $\rmC$ and a corresponding decoder aims to estimate a target by $\hat{\rmZ}$, which may differ from the input source $\rmX$, particularly in scenarios with side information $\rmY$ (see Fig. \ref{fig_system_model}). 
Our primary focus is on the challenge of compressing CSI where a UE encodes observed downlink (DL) CSI into a fixed-length codeword. This compressed representation is then transmitted to a BS, which aims to predict future CSI by leveraging the received codeword along with uplink (UL) CSI as side information. The correlation between UL and DL CSI, stemming from their frequency-invariant characteristics \cite{vasisht2016eliminating, han2023fdd}, makes UL CSI valuable information for improving compression efficiency. 
Additionally, we address the scenario of CSI compression without side information. Across both settings, it is observed that the proposed diffusion model-based compression method significantly outperforms existing approaches in terms of compression effectiveness.
These problems align with ongoing, intensive studies within wireless communication standards \cite{3gpp38214, 9927255}.
Our contributions can be summarized as follows:
\begin{itemize}
    \item We propose a new fixed-rate coding for a computing scheme with side information, leveraging conditional diffusion models. Our method combines efficient vector quantization with trainable codebook-based encoding, where the input source is compressed using neural modules and quantized via the trained codebook. Decoding is achieved through a deterministic backward diffusion process \cite{yang2024lossy}, conditioned on the codeword and side information.
    \item We demonstrate the effectiveness of our diffusion-based coding scheme in various DL CSI compression scenarios in MIMO communications with Clustered Delay Line (CDL) models in NVIDIA Sionna \cite{hoydis2022sionna} and COST2100 \cite{csinet_ver1}. The DL CSI compression is an area of significant interest and active discussion within telecommunications standardization \cite{9927255} due to its key role in improving communication efficiency. 
    \item  The simulation results demonstrate that the proposed method significantly surpasses existing CSI compression techniques. For instance, the proposed scheme requires less than half the data rate of competing methods to achieve the comparable distortion and demonstrates a more than {\emph{twofold improvement}} in some cases. 
    These findings suggest there is room for improvement in the development of the CSI compression techniques. But there may be challenges in the complexity of this approach. We further discuss how to address these challenges.
\end{itemize}

\section{System Model and Problem Formulation}
\label{section:system_model}
We consider the system model depicted in Fig. \ref{fig_system_model}, where the encoder processes an input source $\rmX \in \mathcal{X}$ and compresses it into a codeword $\rmC \in \mathcal{C}$, with the codeword space defined as $\mathcal{C} = \{0,1\}^{B}$, representing a $B$-bit fixed-rate compression. The decoder receives the codeword $\rmC$ along with side information $\rmY \in \mathcal{Y}$. The objective of the decoder is to minimize the distortion between its output $\hat{\rmZ} \in \mathcal{Z}$ and a target $\rmZ \in \mathcal{Z}$. The distortion is measured using a function $d:\mathcal{Z} \times \mathcal{Z} \mapsto \mathbb{R}_{+}$, where $\mathbb{R}_{+}$ represents the space of non-negative real numbers.\\
\indent We employ parameterized models for both the encoder and the decoder, denoted by parameter sets $\theta_{\text{enc}}$ and $\theta_{\text{dec}}$, respectively. The encoder function is represented as $f_{\text{enc}}: \mathcal{X} \mapsto \mathcal{C}$, and the decoder function as $f_{\text{dec}}: \mathcal{C} \times \mathcal{Y} \mapsto \mathcal{Z}$. Given the parameter set $\theta_{\text{enc}}$, the encoder generates a codeword $\rmC = f_{\text{enc}}(\rmX; \theta_{\text{enc}})$. Similarly, the decoder, with parameters $\theta_{\text{dec}}$, estimates the target as $\hat{\rmZ} = f_{\text{dec}}(\rmC, \rmY; \theta_{\text{dec}})$, and $\theta = (\theta_{\text{enc}},\theta_{\text{dec}})$.\\
\indent The system's optimization objective is to minimize the expected distortion as follows.
\begin{alignat}{2}
        & \minimize_{ \theta }
        &\qquad & \mathbb{E}_{p_{\rmX, \rmY, \rmZ}} [ d( f_{\text{dec}}(f_{\text{enc}}(\rmX ;\theta_{\text{enc}}), \rmY; \theta_{\text{dec}}), \rmZ ) ]
        \label{objective_function_01},
\end{alignat}
where $p_{\rmX,\rmY,\rmZ}$ denotes the joint probability distribution of the input source, the side information, and the target. For the distortion measure $d$, we consider the element-wise squared error.

\section{Proposed Approach}
In this section, we introduce a novel fixed-rate coding for computing framework that leverages conditional diffusion models to efficiently compress the input sources and reconstruct target outputs, utilizing available side information. Our encoding scheme employs vector quantization via a trainable codebook for fixed-rate encoding, while the decoding process incorporates a diffusion model conditioned on both the codeword and side information.

\label{section:proposed_approach}
\begin{figure}
\begin{center}
\centerline{\includegraphics[width=1.0\columnwidth]{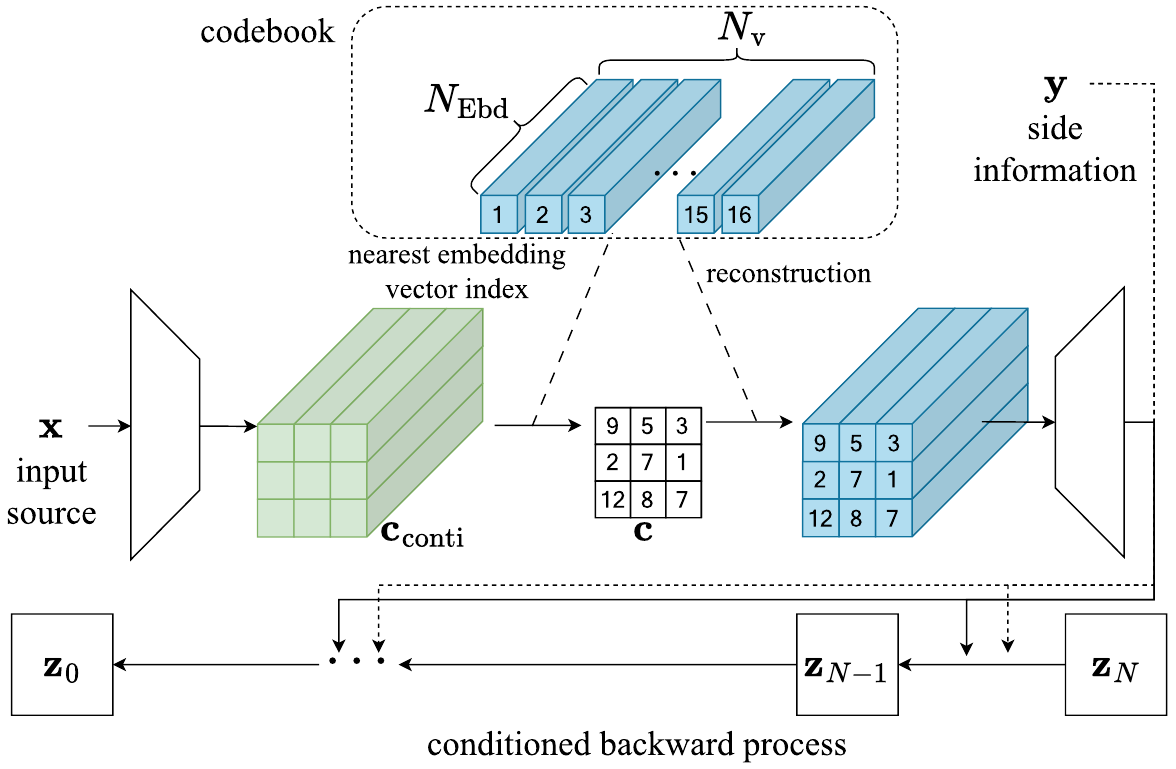}}
\caption{Proposed compression framework.}
\label{fig_model_architecture}
\end{center}
\vskip -0.4in
\end{figure}
\subsection{Fixed-rate encoding with trainable codebook}
As depicted in Fig. \ref{fig_model_architecture}, the encoder takes an input $\rvx$, a realization of $\rmX$, and compresses it into a codeword $\rvc$ in a discrete space. We adopt a discretization module that utilizes a trainable codebook \cite{van2017neural}. Specifically, the input $\rvx$ is processed through neural network layers and transformed into a set of continuous-valued vectors $\rvc_{\text{conti}}$, each with $N_{\text{Ebd}}$ dimensions. The codebook consists of $N_{\text{v}}$ vectors, each of size $N_{\text{Ebd}}$. Each vector in $\rvc_{\text{conti}}$ is replaced by the closest codebook vector based on the minimum $L^2$ distance among the $N_{\text{v}}$ vectors, thereby achieving quantization. The codebook, or set of embedding vectors, is trained alongside the model using the following loss function $L_{\text{cb}}$:
\begin{align}
    \label{codebook_loss}
    L_{\text{cb}} = \Vert \texttt{sg}[\rvc_{\text{conti}}] - \rve \Vert^{2} + \Vert \rvc_{\text{conti}} - \texttt{sg}[\rve] \Vert^{2},
\end{align}
where $\texttt{sg}$ represents the stop-gradient operation, which treats its input as a constant, preventing gradient backpropagation through it. The variable $\rve$ refers to the selected embedding vectors corresponding to $\rvc_{\text{conti}}$. This loss function ensures that the encoder’s output remains close to the selected codebook vectors while simultaneously guiding the codebook vectors to align with the encoder’s output.

\subsection{Conditional diffusion model-based decoding with side information}
In alignment with existing diffusion-based compression methods, we utilize a diffusion backward process to estimate the target at the decoder. Specifically, we adopt the conditional diffusion model proposed in \cite{yang2024lossy} for the decoding process, applying it to a codeword and side information-conditioned denoising diffusion process.

Given the codeword $\rmC\!=\!\rvc$ from the encoder and the available side information $\rmY\!=\!\rvy$, the ultimate goal of the decoder is to sample $\rmZ \sim p(\rvz|\rvc, \rvy)$ via a conditional denoising diffusion process from a $\rvz_{T}$, a realization of $\rmZ_{T}$, where the joint distribution of the target $\rmZ = \rmZ_{0}$ and its noise-perturbed states $\{ \rmZ_{t} \}_{t=1}^{T}$ is modeled as
\begin{align}    \label{conditional_denoising_diffusion_process}
    p(\rvz_{0:T}|\rvc, \rvy) = p(\rvz_{T})\prod \mathcal{N} (\rvz_{t-1} ; \mu_{\theta}(\rvz_{t}, \rvc, \rvy, t), \beta_{t}\mI ).
\end{align}
Here, $\rvz_{0:T}=(\rvz_{0},\ldots, \rvz_{T})$, a realization of $(\rmZ_{0},\ldots, \rmZ_{T})$, and $\mu_{\theta}$ denotes the parameterized mean function, while $\beta_{t}$ represents the variance schedule \cite{song2022denoisingdiffusionimplicitmodels}. The noise-perturbed states evolve according to $q(\rmZ_{t}\vert \rvz_{t-1}) \!=\! \mathcal{N}(\rmZ_{t}; \sqrt{1-\beta_{t}}\rvz_{t-1}, \beta_{t}\mI)$.

Once the mean function $\mu_{\theta}$ is learned, an instance $\rvz$ can be sampled by DDIM sampling \cite{song2022denoisingdiffusionimplicitmodels}. $\mu_{\theta}$ can be obtained by minimizing the upper bound of the negative log-likelihood of the target distribution as follows.
\begin{subequations}
    \label{log_likelihood_upper_bound}
    \begin{align}
        -\log p(\rvz_{0}&|\rvc, \rvy) \le \mathbb{E}_{\rmZ_{1:T} \sim q(\rvz_{1:T}|\rvz_{0})} \Big [ \log \frac{q(\rmZ_{1:T}|\rvz_{0})}{p(\rmZ_{0:T}|\rvc, \rvy)} \Big ] \\
    &\approx \mathbb{E}_{\rmZ_{0},\rmT} \frac{\bar{\alpha}_{\rmT}}{1-\bar{\alpha}_{\rmT}} \Vert \rmZ_{0} - D_{\theta} (\rmZ_{\rvt}, \rvc, \rvy, \rmT)\Vert^2.
    \end{align}
\end{subequations}
The inequality arises from the variational upper bound \cite{ho2020denoising}, with the approximation derived from \cite{salimansprogressive, yang2024lossy}. In this expression, $\rmT$ is a scalar random variable such that $\rmT \sim \mathcal{U}(0,\ldots, T)$, $\bar{\alpha}_{t} = \prod_{j=1}^{t}\alpha_{j}$ with $\alpha_{t} = 1-\beta_{t}$, and $D_{\theta}$ represents the denoising function that takes as input the $t$-step perturbed target $\rvz_{t}$, codeword $\rvc$, side information $\rvy$, and the step number $t$, and outputs the predicted target. From this, we have
\begin{align}
    &\mu_{\theta}(\rvz_{t}, \rvc, \rvy, t) = \frac{1}{\sqrt{\alpha_{t}}}\!\left(\rvz_{t} - {\beta_{t}} \left(  \frac{\rvz_{t} - \sqrt{\bar{\alpha}_{t}} D_{\theta}(\rvz_{t}, \rvc, \rvy, t) }{{1-\bar{\alpha}_{t}}} \right)\!\right).
\end{align}
Based on this, the decoding operation $f_{\text{dec}}(\rvc, \rvy; \theta_{\text{dec}})$ follows an iterative process as expressed in (\ref{decoding_function}), starting from $t\! =\! T$ proceeding sequentially down to $t\! =\! 1$. We initialize the process by setting $\rvz_{T} = \mathbf{0}$.
\begin{align}
    \label{decoding_function}
    \rvz_{t-1} &= \sqrt{\bar{\alpha}_{t-1}} D_{\theta}(\rvz_{t}, \rvc, \rvy, t) \nonumber \\ &+ \sqrt{1-\bar{\alpha}_{t-1}} \left(  \frac{\rvz_{t} - \sqrt{\bar{\alpha}_{t}} D_{\theta}(\rvz_{t}, \rvc, \rvy, t) }{\sqrt{1-\bar{\alpha}_{t}}} \right).
\end{align}

\subsection{Training}
By combining the codebook loss $L_{\text{cb}}$ with the approximated likelihood loss in (\ref{log_likelihood_upper_bound}), we optimize the likelihood function during the training of the codebook. The total loss function to be minimized is defined as
\begin{align}
    \label{loss}
    L = \mathbb{E}_{\rmZ_{0},\rmT} \frac{\bar{\alpha}_{\rmT}}{1-\bar{\alpha}_{\rmT}} \Vert \rmZ_{0} - D_{\theta} (\rmZ_{\rmT}, \rvc, \rvy, \rmT)\Vert^2 + \eta L_{\text{cb}},
\end{align}
where $\eta$ is a hyperparameter that controls the weighting of the codebook loss. The complete training procedure is detailed in Algorithm \ref{algorithm_training}, where the function \texttt{GradientDescent}($\theta$, $L$) represents a single gradient descent update step with respect to the objective function $L$, parameterized by $\theta$. 
\begin{subfigures}
\begin{figure}
\vspace{-0.1in}
\end{figure}
\begin{algorithm}[t]
  \caption{Training }
\label{algorithm_training}
  \begin{algorithmic}[1]
    \Require Initial model $(\theta_{\text{enc}}, \theta_{\text{dec}})$, codebook loss weight factor $\eta$, $\{\bar{\alpha}_{t}\}_{t=0}^{T}$, \texttt{GradientDescent} optimizer
    \vspace{.3em}
    \Ensure Updated $\theta_{\text{enc}}, \theta_{\text{dec}}$
    \vspace{.3em}
    
    \For{$i=0$ {\bfseries to} $N_{\text{train}}$}  
    \vspace{.1em}
    \vspace{.1em}
    \State \parbox[t]{\dimexpr\linewidth-\algorithmicindent}{Sample $(\rvx,\rvy,\rvz)\sim p_{\rmX, \rmY, \rmZ}$, $t \sim \mathcal{U}(0,\ldots, T)$, $\epsilon \sim \mathcal{N}(0, \mI)$
    }

    \State $\rvc = f_{\text{enc}}(\rvx; \theta_{\text{enc}})$ 

    \State $L_{\text{cb}} = \Vert \texttt{sg}[\rvc_{\text{conti}}] - \rve \Vert^{2} + \Vert \rvc_{\text{conti}} - \texttt{sg}[\rve] \Vert^{2}$

    \State \parbox[t]{\dimexpr\linewidth-\algorithmicindent}{%
    $L = \frac{\bar{\alpha}_{t}}{1-\bar{\alpha}_{t}} \Vert \rvz_{0} - D_{\theta} (\sqrt{\bar{\alpha}_{t}} \rvz_{0} + \sqrt{1-\bar{\alpha}_{t}} \epsilon, \rvc, \rvy, t)\Vert^2 + \eta L_{\text{cb}}$%
}

    \State \texttt{GradientDescent}($\theta$, $L$)
\EndFor

  \end{algorithmic}
\end{algorithm}
\end{subfigures}

\subsection{Model Architecture}
\label{appendix_model_architecture}
For the encoder and decoder design, we adopt the neural layers described in \cite{yang2024lossy}, including Downsampling Units (DU), Upsampling Units (UU), ResNet Blocks (RNB), Linear Attention Layers (Attn), a Convolutional Layers (Conv), and a Transposed Convolution (ConvT). The following model description is based on an input source $\rmX$ that has been preprocessed into a $32 \! \times \! 32 \! \times \! 2$ tensor (see Section \ref{subsubsection_simulation_settings_1} for preprocessing details).

\emph{Encoder.}
The encoder comprises two RNB + DU blocks, which increase the channel dimensions while reducing the spatial dimensions of the input tensor. These blocks progressively raise the channel dimensions to 64 and then to 128, producing an output tensor of $8 \! \times \! 8 \! \times \! 128$ for quantization.

\begin{figure*}[t]
\begin{center}
\centerline{\includegraphics[width=0.70\linewidth]{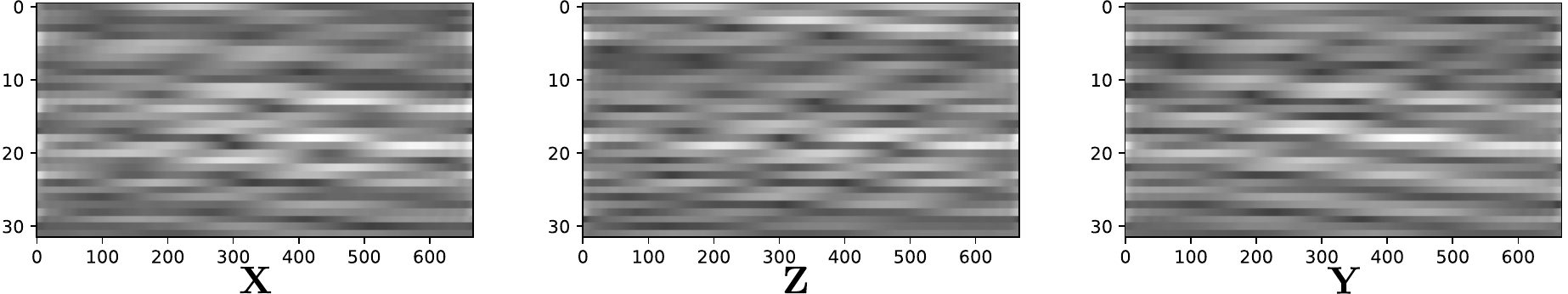}}
\caption{Magnitude visualization of CSI samples: The task is to predict $\rmZ$ (future DL CSI) from compressed $\rmX$ (current DL CSI), leveraging $\rmY$ (UL CSI) as side information.}
\label{csi_samples_sf_domain}
\end{center}
\vspace{-0.4in}
\end{figure*}

\emph{Quantization.}
The encoder output undergoes quantization via a codebook-based vector quantization scheme. The output tensor, structured as 64 vectors (from the $8 \times 8$ layout), each with dimensionality 128, is quantized by replacing each vector with the closest of $N_{\text{v}}$ embedding vectors in the codebook (each of size $N_{\text{Ebd}} = 128$), based on the minimum $L^2$ distance. The selected codebook indices are then concatenated to form the codeword, which is transmitted to the decoder.

\emph{Decoder.}
The decoding process consists of two primary components: (1) reconstructing the quantized latent space from the codeword and (2) executing the diffusion backward process, conditioned on the reconstructed codeword and side information. The decoder begins by receiving the codeword, which contains the vector indices. Using the codebook, it reconstructs the tensor of size $8 \times 8 \times 128$ by selecting embedding vectors corresponding to the codeword indices. This tensor is then processed by two sets of RNB + UU, which progressively upsample and reduce the channel dimensions. After the first RNB + UU, the output size becomes $16 \times 16 \times 64$. A second set of RNB + UU further upsamples the data, resulting in a final tensor of size $32 \times 32 \times 8$. These outputs are used as conditioning information for the diffusion backward process, implemented through a U-Net architecture. The U-Net architecture leverages the output of the decoder and additional side information (if available) as conditioning inputs. The side information is concatenated with the decoder's last output with size $32 \times 32 \times 8$, and the time step information is also provided as input to the U-Net. For the U-Net layers, we use an embedding dimension of 64 and dimension multipliers of 1, 2, 3, and 4 for the downsampling and upsampling stages, following the architecture outlined in \cite{yang2024lossy}. If no side information is provided, the concatenation step is omitted, and the U-Net processes only the decoder output.

\section{Numerical Results}
\label{section_experiment}
\subsection{Hyperparameters and Performance Metric}
We set $\eta=4.5\times 10^{-4}$ as the weight in the codebook loss in (\ref{loss}). The model is trained for a total of $N_{\text{train}}\!=\!3 \times 10^5$ steps, using a batch size of 100. We utilize the Adam optimizer \cite{kingma2014adam} for $\texttt{GradientDescent}$ training, with a learning rate of $10^{-3}$.
For the diffusion backward process, we set the number of denoising steps $T=4$ and apply a cosine variance schedule, with corresponding values for the noise parameters: $\alpha_{1}=8.47 \times 10^{-1}$, $\alpha_{2}=4.93 \times 10^{-1}$, $\alpha_{3}=1.44 \times 10^{-1}$, and $\alpha_{4}=1.44 \times 10^{-4}$.

The performance is evaluated using the Normalized Mean Squared Error (NMSE), defined as $\mathbb{E}[\Vert \rvz - \hat{\rvz} \Vert^2_{\text{F}} / \Vert \rvz\Vert^2_{\text{F}}]$, where $\Vert \cdot \Vert_{\text{F}}$ denotes the Frobenius norm of the matrix.

\subsection{CSI Compression for channel prediction}
\label{subsection_csi_prediction}
\subsubsection{Simulation Settings}
\label{subsubsection_simulation_settings_1}
We utilize NVIDIA Sionna \cite{hoydis2022sionna} for real-time CSI data generation. 
We simulate the CSI of a MIMO system in a downlink configuration, where the BS is equipped with 32 antennas and the UE is equipped with a single antenna.
The system operates in the frequency domain with 667 Orthogonal Frequency Division Multiplexing (OFDM) subcarriers, each spaced 15 kHz apart. The CDL-C profile simulates the CSI with a delay spread of 300 ns. The carrier frequency is set to 2.11 GHz for the downlink and 1.91 GHz for the uplink. Channel variations over time are simulated at an interval of 5 ms to generate future CSI instances. To model realistic mobility, the UE is simulated at a speed of 5 m/s, reflecting typical vehicular scenarios.
The CDL model is parameterized using an omnidirectional antenna pattern at the UE and the 3rd Generation Partnership Project (3GPP) technical specification (TS) 38.901 antenna pattern at the BS \cite{3gpp.38.901}. Both the BS and UE antennas are vertically polarized. 

The setup results in a $32 \times 667$ complex matrix for the DL CSI, UL CSI, and the future DL CSI (target). We simulate the channel's time evolution by generating 71 consecutive time slots (14 OFDM symbols per 5 slots, plus one), with the first slot serving as the input source $\rmX$ and the last slot's UL and DL information serving as the side information $\rmY$ and the target $\rmZ$ for future DL CSI prediction, respectively. This UL CSI is correlated with the DL CSI through frequency-invariant characteristics \cite{vasisht2016eliminating, han2023fdd}. 
We assume perfect UL CSI acquisition, and the prediction is performed within the same time slot. For simplicity, each CSI instance contains only a single time slot information instead of the full 14 time slots. 
A sample of the input, target output, and side information is provided in Figure \ref{csi_samples_sf_domain}.

To reduce computational overhead, we first apply the 2D Inverse Fast Fourier Transform (IFFT) to the complex matrices $\rmX, \rmY$, and $\rmZ$, converting the data from the spatial-frequency domain to the angular-delay domain. This transformation induces sparsity in the data, as supported by certain assumptions \cite{wang2018spatial}. We then retain only the first 32 elements in the delay domain, as the remaining values tend to zero, yielding cropped angular-delay domain representations for $\rmX$, $\rmZ$, and $\rmY$. The original CSI can be reconstructed by appending zero matrices of size $32 \times 635$ and performing a 2D FFT. The preprocessing is a widely adopted technique for efficient CSI representation \cite{csinet_ver1, csinet_ver2, lu2020crnet}, and we evaluate the NMSE performance within the cropped angular-delay domain.

\subsubsection{Baselines}
To evaluate our proposed method, we benchmark it against the CsiNet \cite{csinet_ver1} and CRNet \cite{lu2020crnet} models. Initially designed for CSI compression without accounting for side information or bit-level quantization, these baselines were minimally adapted to enable a fair comparison.

(a) \emph{CsiNet with Uniform Quantization.} We employ the encoder and decoder architectures from \cite{csinet_ver1}, training the model with Mean Squared Error (MSE) loss between the network output and the target representation, $\rmZ$. In the original CsiNet, the latent representation is a continuous vector, denoted $\rvc_{\text{conti}}$ whose dimension is $N_{\text{cl,f}}$. To achieve discrete codeword, we apply 6-bit uniform quantization to each element of the latent vector $\rvc_{\text{conti}}$. Specifically, the encoder’s output is constrained within [-1, 1] by a hyperbolic tangent activation ($\texttt{tanh}$). Each element of the latent vector is quantized with 6 bits, resulting in a $(N_{\text{cl,f}} \times 6)$-bit length codeword.
To ensure gradient-based optimization remains effective, we use a stop-gradient approach, updating encoder parameters based on gradients from the pre-quantized values as
$\rvc = \rvc_{\text{conti}} + \texttt{sg}[\rvc - \rvc_{\text{conti}}] $ where $\rvc$ is the discretized input to the decoder. This design supports end-to-end training of the encoder and decoder. Side information is integrated into decoding by modifying the original $\texttt{ResidualBlock}$ in \cite{csinet_ver1}, where the side information is concatenated with the input being refined.
For consistency, input data is scaled to lie within [-1, 1], and we apply $\texttt{tanh}$ to the decoder output.

(b) \emph{CRNet with Uniform Quantization.} Similarly, we adapt the encoder and decoder architectures from \cite{lu2020crnet}, which also use a floating-point latent vector. The quantization follows the same procedure as for CsiNet, applying 6-bit quantization after constraining values to [-1, 1] via a $\texttt{tanh}$ activation function.
To incorporate side information, we concatenate it with the decoder’s input. Specifically, the CRNet decoder first takes the quantized codeword $\rvc$ of size $N_{\text{cl,f}}$, applies a dense layer to match the target dimensionality, and reshapes it to the target form of $32\!\times \!32 \! \times \! 2$. The side information is then concatenated, forming a tensor of size $32\!\times \!32 \! \times \! 4$ for further decoding.
The parameters in this neural network are trained using MSE loss to minimize reconstruction distortion with respect to the target.

\begin{figure}[t]
\includegraphics[width=0.95\columnwidth]{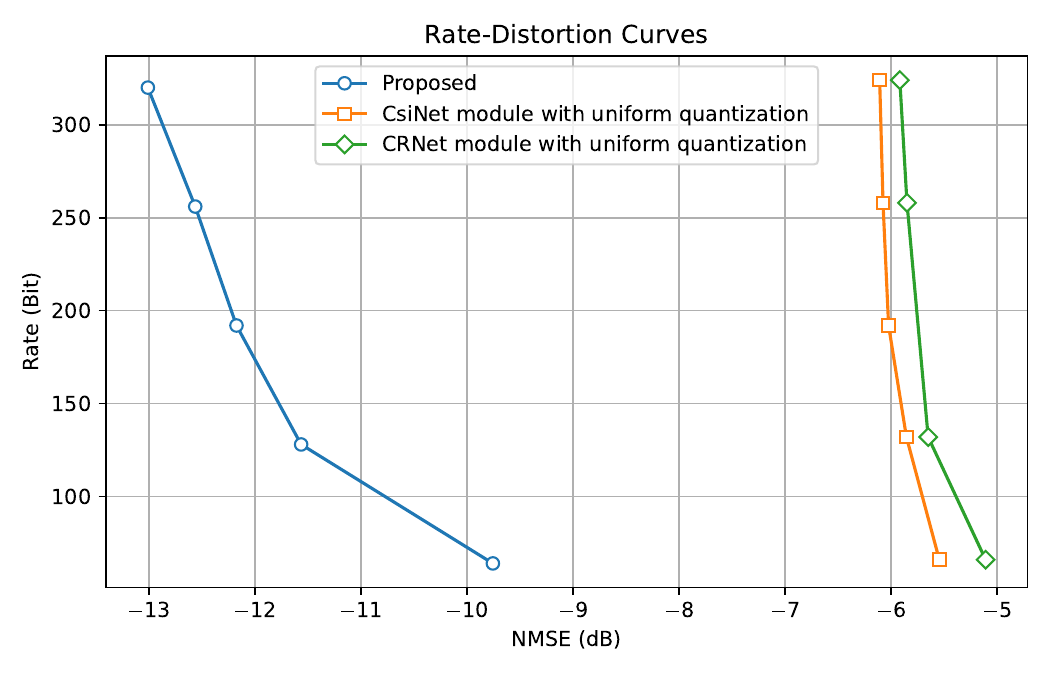}
\caption{Rate-distortion curves for experiments in Sec. \ref{subsection_csi_prediction}.}
\label{rd_curves_cdl}
\vspace{-0.2in}
\end{figure}

\subsubsection{Results}
The rate-distortion curves comparing the proposed method with the baseline models are presented in Fig. \ref{rd_curves_cdl}. The results demonstrate that the proposed method outperforms the baseline methods, as the proposed method achieves lower distortion for a given rate. In contrast, the baseline methods exhibit performance saturation around -6 dB NMSE, where increasing the bit rate provides only marginal improvements. For instance, increasing the bit rate by over 100 bits at 132-bit rate yields less than a 0.5 dB gain, likely due to the limited representational capacity of the neural modules or inefficient encoding schemes. In contrast, the proposed method demonstrates a gain exceeding 1 dB with an increase of just 64 bits (from 64-bit to 128-bit compression) and achieves an additional gain of over 0.5 dB when progressing from 128-bit to 192-bit compression. These results suggest that straightforward extensions of existing neural lossy compression methods may be suboptimal for CSI compression tasks, particularly in the context of future CSI prediction.

\subsection{CSI Compression for channel reconstruction}
\label{subsection_csi_compression}
\subsubsection{Simulation Settings}
To further assess the proposed compression approach, we conduct experiments focused on CSI compression aiming to reconstruct $\rmX$. This task serves as a simplified version of the prediction scenario described in Section \ref{subsection_csi_prediction}, with the objective limited to reconstructing the input source $\rmX$ (i.e., $\rmZ\!=\! \rmX$) without incorporating side information. For simulation, we use the widely adopted COST2100 outdoor dataset \cite{csinet_ver1}, comprising $10^{5}$ training samples and $2\! \times \! 10^{4}$ test samples.

\subsubsection{Baselines}
We compare the performance of our fixed-rate coding scheme against the following state-of-the-art CSI compression algorithms: (a) CsiConformer \cite{sun2024efficient}, which integrates convolutional operations with self-attention mechanisms to enhance CSI feedback accuracy. In Fig. \ref{rd_curves_cost2100}, we plot the performance of this approach with the relevant quantization methods reported in the work \cite{sun2024efficient}, (b) CsiNet+ \cite{guo2020csinetplus}, which improves performance by fine-tuning the decoder parameters based on quantization bits. The authors introduce an offset network to mitigate the impact of quantization distortion, thereby improving overall compression performance, (c) Deep AE Entropy Coding \cite{ravula2021deep}, where the authors propose entropy-based coding for CSI quantization. Note that entropy coding allows variable-length codewords depending on the input instance, and the rate-distortion curve of entropy coding may act as a lower bound for fixed-rate coding in ideal scenarios. 

\subsubsection{Results}
\begin{figure}
\begin{center}
\centerline{\includegraphics[width=0.95\columnwidth]{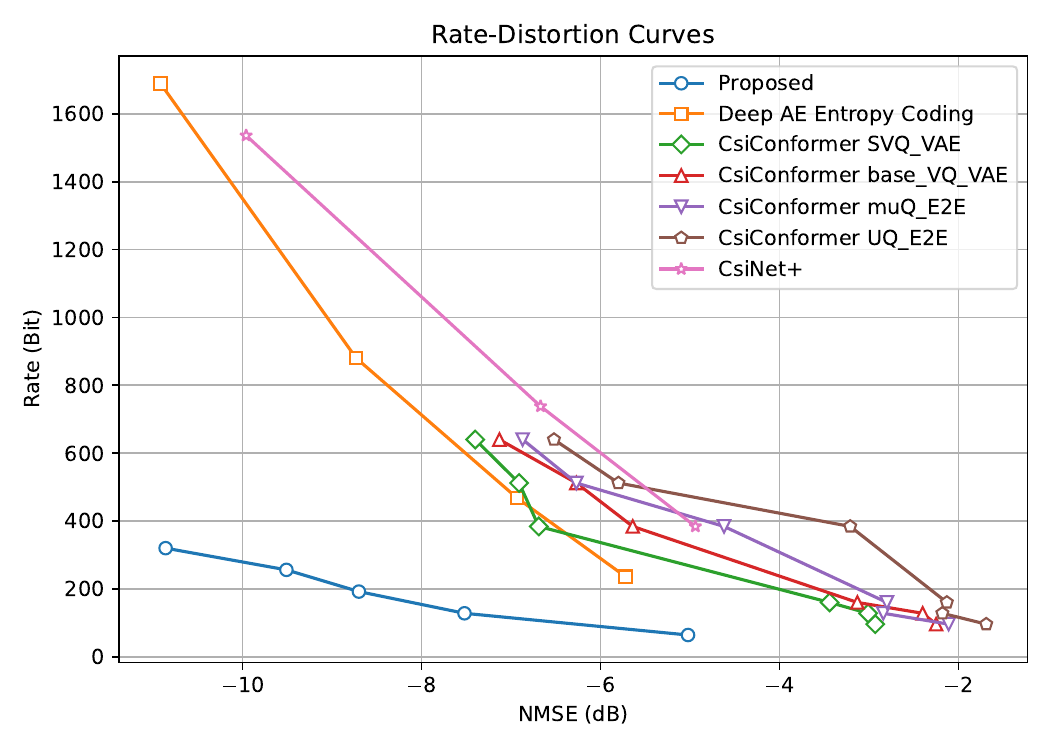}}
\caption{Rate-distortion curves for experiments in Sec. \ref{subsection_csi_compression}.}
\label{rd_curves_cost2100}
\end{center}
\vspace{-0.4in}
\end{figure}
In Fig. \ref{rd_curves_cost2100}, the rate-distortion curves for the baseline methods and the proposed method are presented. The baseline performance data are sourced directly from \cite{sun2024efficient, guo2020csinetplus, ravula2021deep}, where the performance of the baseline CsiConformer model is reported with the relevant quantization methods. The results show that the proposed scheme significantly outperforms the existing approaches, as its rate-distortion curve consistently forms the lower bound compared to the other baselines. For instance, to achieve a distortion of approximately -7 dB, the proposed method requires fewer than 150 bits, whereas all baseline methods require more than 400 bits, demonstrating suboptimal performance. Notably, the proposed fixed-rate coding scheme also outperforms deep autoencoder-based entropy coding, which allows variable-length codewords based on the input instance. This highlights the superior efficiency of the proposed method, showing that it can significantly reduce the number of bits required for a given distortion level, thus offering substantial savings in radio resources for communication.

\section{Discussion}
We presented a new fixed-rate coding scheme with side information for DL CSI compression, which uses a vector quantization method using a trainable codebook and the diffusion-based backward process for decoding, conditioned on both the codeword and side information. Experimental results demonstrated that the proposed method significantly outperformed state-of-the-art CSI compression techniques, effectively reducing the required bit rate for a given distortion across diverse scenarios. These findings highlight the strong potential of our scheme for future network applications, where minimizing transmission rates is crucial. Further research could focus on enhancing computational efficiency by exploring more lightweight architectures \cite{ruan2022malunet} or fast sampling processes \cite{liuflow}, increasing the practicality of the scheme for real-world deployment.

\bibliographystyle{IEEEtran}
\bibliography{references_diffusion, references_csi_feedback, references_neural_compression, references_optimization}
\end{document}